\documentstyle[prc,aps,epsfig,multicol]{revtex}
\draft
\tightenlines
\newcommand{\ber}{\begin{eqnarray}}
\newcommand{\eer}{\end{eqnarray}}
\begin{document}
\title{Strangeness equilibration in heavy ion collisions}
\author{Subrata Pal, C.M. Ko, and Zi-wei Lin}
\address{Cyclotron Institute and Physics Department,
Texas A\&M University, College Station, Texas 77843-3366}

\maketitle

\begin{abstract}
Using a relativistic transport model for heavy ion collisions at 
energies that are below the threshold for kaon and antikaon
production in nucleon-nucleon collisions, we study how their
abundances approach the canonical equilibrium during 
the collisions. 
We find that kaons are far from chemical equilibrium at 
the initial and high density stage, 
and they approach equilibrium only during the expansion stage of
the collisions when their production rate is small 
and becomes comparable to their annihilation rate.   
In contrast, antikaons approach chemical 
equilibrium much earlier but eventually fall out of equilibrium
again as a result of their large 
annihilation cross sections in nuclear matter.

\end{abstract}

\begin{multicols}{2}

Recent analyses have shown that most hadrons measured in heavy ion 
collisions can be described by statistical models
based on the grand canonical ensemble for abundant particles
\cite{bra99} and the canonical ensemble for rare particles 
\cite{red}. On the other hand, studies based on transport model 
indicate that the chemical equilibration time for kaons and antikaons 
in the hot dense matter, that is expected to be formed in heavy ion 
collisions at the AGS and SPS energies, is an order of magnitude 
longer than the heavy ion collision time \cite{brat}. It has thus 
been suggested \cite{brown} that the kaon equilibration time can be 
significantly shortened if the kaon mass is reduced in dense matter 
as a result of the large attractive scalar interaction and diminishing 
repulsive vector interaction due to the assumption of vector decoupling. 
The disappearance of the repulsive vector interaction is, however,
not consistent with the strong repulsive vector potential that is
needed to understand the vanishing kaon flow observed in heavy ion 
collisions at GSI energies \cite{fopi,liko} and the large kaon 
antiflow observed in heavy ion collisions at AGS energies \cite{e895,pal}. 
Also, a recent transport model study has shown that the measured kaon 
yield in heavy ion collisions at GSI \cite{kaos} can be explained 
with kaons interacting with both the scalar and vector potentials 
\cite{fuchs}. 

To gain insight into this problem, a kinetic theory for the time evolution 
of particle production \cite{rate} can be applied. 
It is found that the equilibrium time, $\tau_0^{\rm C}$, for rare 
particles carrying $U(1)$ charge and described by the canonical ensemble 
due to $U(1)$ charge conservation, is much shorter than what is expected 
from the grand canonical ensemble, i.e.,
\begin{equation} 
\tau_0^{\rm C}=\tau_0^{\rm GC} N_{\rm eq}^{\rm GC} \ll \tau_0^{\rm GC},
\end{equation}
where $N_{\rm eq}^{\rm GC}$ is the average multiplicity of rare particles 
per event if they were described by the grand canonical ensemble. Also, it is
shown that 
\begin{equation}
N_{\rm eq}^{\rm C}=(N_{\rm eq}^{\rm GC})^2 \ll N_{\rm eq}^{\rm GC},
\end{equation}
so the equilibrium multiplicity in the canonical ensemble is much
lower than that given by the grand canonical ensemble.  

The above idea based on the kinetic theory can be quantitatively 
studied using the transport models, which have been very successful 
in understanding many aspects of heavy ion collisions \cite{transport}.
In this paper, we shall use the transport model to study kaon and antikaon
production in heavy ion collisions at energies that are below the
thresholds for their production in nucleon-nucleon interactions,
and to understand how kaons and antikaons approach chemical equilibrium 
in these collisions. In previous studies of subthreshold
kaon production using transport models \cite{kaon,kaon1}, only the production 
of kaons has been included while their annihilation via the inverse reactions
has been neglected. The neglect of kaon annihilation would not be
a bad approximation if kaons are far from chemical equilibrium. 
However, to understand kaon chemical equilibration in transport models,
we need to include kaon annihilation to see how the kaon production and 
annihilation rates become comparable during the finite heavy ion collision
time.

We follow essentially the same model used in Ref. \cite{llb},
which is based on the relativistic transport model RVUU \cite{rvuu}.
In this model, the nuclear potential is taken from the nonlinear 
Walecka model, so it has both an attractive scalar and a repulsive 
vector part. The attractive scalar potential allows one to treat
consistently the change of nucleon mass in nuclear matter. In dense 
matter, the nucleon mass is reduced and the energy is in the scalar
field. As the system expands, the nucleon regain its mass from the
scalar field energy. The model in Ref. \cite{llb} includes kaon 
production from both baryon-baryon and meson-baryon interactions. 
For kaon production from baryon-baryon interactions, the cross sections 
are taken from the predictions of the meson-exchange model introduced in
Ref. \cite{kcross}. For kaon production cross sections from 
meson-baryon interactions, they are obtained from the resonance model 
of Ref. \cite{tsushima}. The produced kaons together with their partners, 
mainly hyperons, not only undergo elastic scatterings with
baryons but are also affected by mean-field potentials. For kaon
and hyperon scattering cross sections, we take the empirical values 
as in Ref. \cite{llb}. 
The kaon potential is taken from the chiral Lagrangian including both
scalar and vector interactions \cite{kaplan} with their strengths 
determined from experimental observables such as the kaon yield and
collective flow in heavy ion collisions \cite{llb}. 

Since the kaon production probability is much smaller than 
one in heavy ion collisions at subthreshold energies, it is usually
treated by the perturbative method, i.e., the effect of kaon production
on the dynamics of heavy ion collisions is neglected.
Specifically, a kaon is produced in a baryon-baryon
or a meson-baryon interaction that is above the production 
threshold. The produced kaon then carries a probability given
by the ratio of the kaon production cross section to the 
total baryon-baryon or meson-baryon cross section. Furthermore, 
to treat the rescattering of kaons and hyperons with other 
particles, one allows more than one pair of kaon and hyperon to be 
produced in such a collision. When these kaons and hyperons 
are scattered by other particles, their momenta are changed
according to the differential cross section, which is usually taken to 
be isotropic, while changes of the momenta of the other particles
are neglected. The final kaon yield and spectrum are then obtained
by adding the probabilities of individual kaons and dividing by the
number of kaons produced in each baryon-baryon or meson-baryon
collision. 

To study the chemical equilibration of kaons, we improve the model of Ref.
\cite{llb} by including also kaon annihilation by hyperon, using 
the cross sections determined from the meson-baryon interactions 
via the detailed balance relations. Because of strangeness conservation, 
a kaon is produced together with a hyperon. As the kaon production 
probability is much less than one in heavy ion collisions at 
subthreshold energies, there is only one hyperon in an event in 
which a kaon is produced. Kaon annihilation can thus occur only
when there is a collision between the same pair of kaon and hyperon 
that is produced in the baryon-baryon or meson-baryon interaction.
As shown in Ref. \cite{rate} via the kinetic equation, the annihilation
between such a pair of particles that are produced simultaneously 
as a result of the $U(1)$ charge conservation would lead to an 
equilibration described by the canonical ensemble. 

\begin{figure}[ht]
\centerline{\epsfig{file=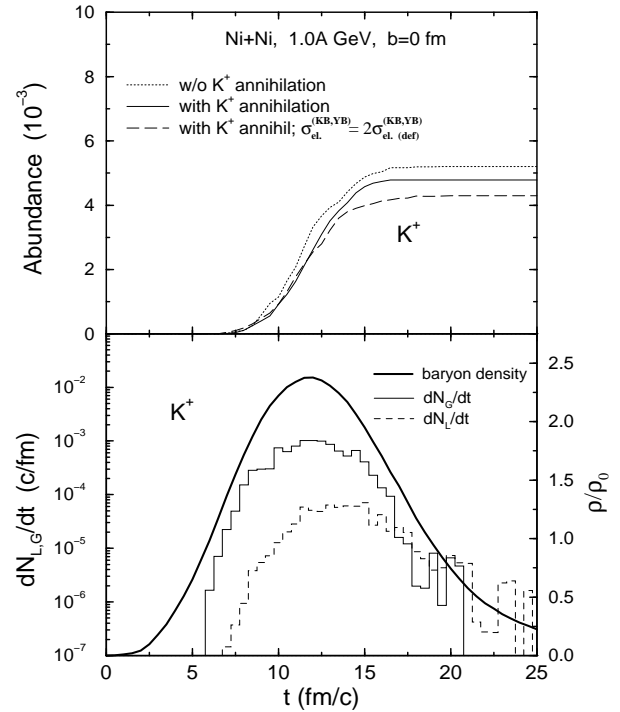,width=3in,height=3.7in,angle=0}}
\vspace{0.1cm}

\caption{
Top panel: Time evolution of kaon abundance in Ni+Ni collisions
at $1A$ GeV and impact parameter $b=0$. Dotted curve is the result 
without kaon absorption while solid and dashed curves are results with 
kaon absorption from using default $KB$ and $YB$ elastic scattering  
cross sections and twice the cross sections, respectively. Bottom panel:
Time evolution of kaon production rate (solid curve) and absorption
rate (dashed curve) as well as the central baryon density (thick solid
curve).}
\label{kaon}
\end{figure}

To illustrate the physics of kaon chemical equilibration in heavy
ion collisions, we consider Ni+Ni collisions at $1A$ GeV 
and impact parameter $b=0$ fm, which is below the threshold for
both kaon and antikaon production in nucleon-nucleon interactions.
Similar results are obtained for collision energies at $2A$ GeV,
which is above the kaon production threshold but below the antikaon
production threshold in nucleon-nucleon interactions. In the top panel 
of Fig. \ref{kaon}, the time evolution of the kaon abundance 
for the scenarios with (solid curve) and without (dotted curve) 
kaon annihilation are given. It is seen that including kaon 
annihilation by hyperon reduces the final kaon yield by only about 10\%. 
As shown in the lower window of Fig. \ref{kaon}, the kaon production
rate (solid curve) is appreciable only when the nuclear density 
(thick solid curve) is high. The effect due to kaon annihilation
is better illustrated by the kaon annihilation rate (dashed curve)
shown in the bottom panel of Fig. \ref{kaon}. We see that
the annihilation rate is negligible during the high density stage
when most kaons are produced. This result thus justifies the 
neglect of kaon annihilation in previous studies. Although the
kaon annihilation rate is small, it becomes comparable to the
kaon production rate at about 17 fm/$c$, indicating that the kaon
yield eventually approaches chemical equilibrium. We note that
the baryon density at which kaons reach chemical equilibrium
is about 1.2 $\rho_0$, where $\rho_0\approx 0.16$ fm$^{-3}$ is
the normal nuclear matter density.
 
\begin{figure}[ht]
\centerline{\epsfig{file=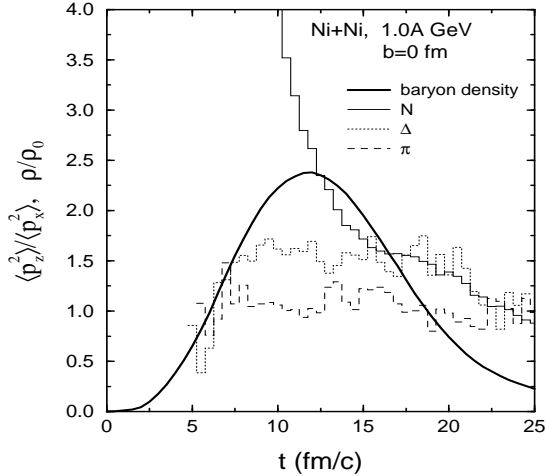,width=2.8in,height=2.5in,angle=0}}
\vspace{0.1cm}

\caption{Time evolution of the ratio 
$\langle p_z^2 \rangle/\langle p_x^2 \rangle$ 
for nucleons (solid curve), deltas (dotted curve), 
and pions (dashed curve). The central baryon density as a function of 
time is given by the thick solid curve.}
\label{thermal}
\end{figure}

To better understand kaon production in heavy ion collisions, 
it is useful to know if the system is close to thermal equilibrium 
when kaons are produced. For this purpose, we have evaluated the
average squared momentum of all particles both in
the beam direction, $\langle p_z^2 \rangle$, and in the transverse direction, 
$\langle p_x^2 \rangle$. 
The ratio $\langle p_z^2 \rangle/\langle p_x^2 \rangle$ can
then be used to characterize the degree of thermal equilibrium, with 
a value of one corresponding to complete thermal equilibrium.
In Fig. \ref{thermal}, we show the time evolution of this ratio 
for nucleons (solid curve), deltas (dotted curve), and pions (dashed curve)
in the central volume of 8 fm$^3$. It is seen that at the time
when most kaons are produced, both pions and deltas are close to
thermal equilibrium while nucleons are not. Since kaons are dominantly 
produced from pion and nucleon interactions (about 70\%), our results 
thus show that they are mostly produced from a dense but not 
thermally equilibrated hadronic matter. 

According to the kinetic theory \cite{rate}, the chemical equilibration
time in the canonical formalism is given by $\tau_0^C=V/L=N_K/(dN_L/dt)$,
where $V$ is the volume of the region where kaon annihilation occurs,
$L$ is the momentum averaged cross section for kaon annihilation,
and $N_K$ and $dN_L/dt$ are the kaon number and its absorption rate,
respectively. At time $t=12$ fm/$c$ when the kaon absorption rate is largest, 
we have from Fig. \ref{kaon}, $N_K=4.2\times 10^{-3}$
and $dN_L/dt=7.0\times 10^{-5}$, which give a kaon chemical equilibration
time of 60 fm/c if the system is prevented from expanding from $t=12$ fm/$c$. 
On the other hand, 
the thermal average of the kaon annihilation cross section 
$\sigma_{KY\to\pi N}$ is about 0.25 fm$^2$ at temperature $T=75$ MeV, 
and it changes by less than 20\% for $50<T<100$ MeV due to the exothermic 
nature of the annihilation process. The above value for the
kaon chemical equilibration time implies that the effective volume in which
kaon annihilation occurs is about 15 fm$^3$. Since the chemical 
equilibration time for kaons is much longer than the heavy ion 
collision time, we would normally expect them not to reach chemical 
equilibrium during the collisions. 
However, since the kaon production rate 
decreases strongly as the temperature decreases due to the large 
threshold of the production process, 
its value can thus become comparable to the annihilation rate when
the system expands and cools. When this happens 
for later stage of heavy ion collisions as shown in 
the transport model results of Fig. \ref{kaon},  
kaons can then reach chemical equilibrium. 
We thus note that, since most kaons are produced in the 
dense but non-equilibrium stage, the chemical equilibration time 
could be overestimated when determined in the normal way. 
Furthermore, the chemical freeze-out temperature extracted 
from the final kaon abundance in thermal model analysis 
does not provide sufficient information on the dynamics of kaon production.

The effect due to kaon annihilation depends on the magnitude of
kaon and hyperon elastic scattering cross sections with other particles.
If there are no such scatterings, e.g, if these cross sections are 
set to zero, then the produced kaon and hyperon would simply 
move away from each other without further interactions, leading
to a result similar to that without kaon annihilation. On the
other hand, larger kaon and hyperon scattering cross sections
with other particles would force them into a smaller region,
thus increasing the kaon annihilation rate. This is demonstrated
in the top panel of Fig. \ref{kaon} by the dashed curve, which
is obtained by taking the kaon and hyperon scattering cross sections
with other baryons to be twice the default values. It is seen that
these larger cross sections indeed further reduce the kaon yield.   

\begin{figure}[ht]
\centerline{\epsfig{file=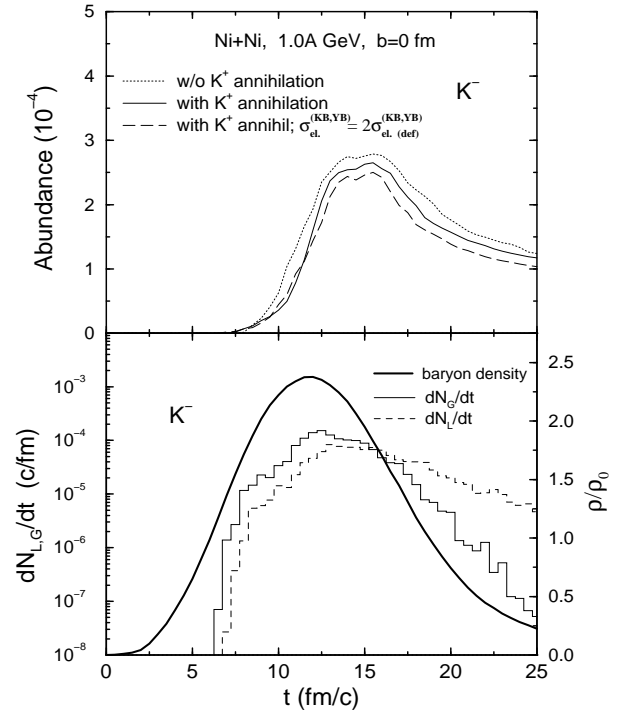,width=3in,height=3.7in,angle=0}}
\vspace{0.1cm}

\caption{Same as Fig. \protect\ref{kaon} for antikaons.}
\label{akaon}
\end{figure}

We also study antikaon production within the relativistic transport model,  
where antikaons are produced not only from 
baryon-baryon and meson-baryon interactions 
but also from meson-hyperon interactions.
The cross sections for antikaon production from both baryon-baryon
and meson-baryon interactions are taken from an analysis based
on the meson-exchange model \cite{sibirtsev}. For antikaon
production from the meson-hyperon interactions, the cross sections
are obtained 
from the empirical cross sections for $K^-$ absorption
by nucleon to form a pion and a hyperon \cite{cugnon}. As first
pointed out in Ref. \cite{ko1}, antikaon production in heavy ion
collisions at subthreshold energies is mainly due to the meson-hyperon
interactions. For antikaon annihilation, it is dominated by 
the reaction ${\bar K}N\to\pi Y$, and this has already been included 
in previous transport model studies \cite{llb,akaon,cassing}.
We note that to account for the observed enhancement of antikaon
production in heavy ion collisions at subthreshold energies, it has been
shown in Refs. \cite{llb,akaon,cassing} that a dropping of antikaon
mass due to medium effects is needed.

The results for antikaon production in Ni+Ni collisions at $1A$ GeV 
and impact parameter $b=0$ fm are shown in Fig. \ref{akaon}. In the top
panel, the time evolution of antikaon abundance is shown for the cases
with (solid curve for default kaon and hyperon elastic scattering
cross sections and dashed curve for twice the default cross sections) 
and without (dotted curve) kaon annihilation. 
The reduction of antikaon yield when kaon annihilation is allowed is
due to the reduction in the production probability of hyperons, which
contribute most to antikaon production. The time evolution of 
the antikaon production and annihilation rates are shown in the 
bottom panel of Fig. \ref{akaon} by the solid and dashed curves, 
respectively. Similar to kaon production, we find that antikaons are 
mostly produced in the high density stage of heavy ion collisions.  
Because of the large baryon density, their annihilation rate
through the reaction ${\bar K}N\to\pi Y$ is larger than the kaon 
annihilation rate. As a result, antikaons approach chemical 
equilibrium even in the earlier stage of heavy ion collisions. 
However, they eventually fall out of equilibrium 
as shown in the lower panel of Fig. \ref{akaon}.

To summarize, we have improved the perturbative treatment of strange
particle production in the relativistic transport model by including
the annihilation of kaon and hyperon that are produced as a pair 
due to strangeness conservation. For heavy ion collisions at
energies below the threshold for strange particle production 
in nucleon-nucleon interactions, we find that both kaons and antikaons 
are largely produced during the high density stage of the collisions
when the system has not reached thermal equilibrium. Because of their large 
annihilation cross sections in dense nuclear matter, antikaons 
are near chemical equilibrium much earlier. For kaons, their 
abundance at high density stage is far from equilibrium, and it only 
becomes close to the equilibrium value during the expansion stage of
heavy ion collisions when the production rate is small and 
comparable to the annihilation rate. The small kaon annihilation rate 
in heavy ion collisions thus justifies the neglect 
of kaon annihilation in previous transport model studies of 
subthreshold kaon production. 

\section*{acknowledgment}

This work was supported by the National Science Foundation under Grant 
No. PHY-9870038, the Welch Foundation under Grant No. A-1358, and the 
Texas Advanced Research Program under Grant No. FY99-010366-0081.

\end{multicols}

\begin{thebibliography}{99}

\bibitem{bra99}
P. Braun-Munzinger, I. Heppe, and J. Stachel,
Phys. Lett. B {\bf 465}, 15 (1999).
\bibitem{red}J. S. Hamieh, K. Redlich, and A. Tounsi,
Phys. Lett. B {\bf 486}, 61 (2000).
\bibitem{brat}
E.L. Bratkovskaya {\it et al.}, 
Nucl. Phys. {\bf A675}, 661 (2000).
\bibitem{brown}
G.E. Brown, M. Rho, and C. Song, nucl-th/001008. 
\bibitem{fopi}J. Ritman, FOPI collaboration, Z. Phys. A {\bf 352}, 355 (1995).
\bibitem{liko}G.Q. Li, C.M. Ko, and B.A. Li, Phys. Rev. Lett. {\bf 74}, 235
(1995); G.Q. Li and C.M. Ko, Nucl. Phys. {\bf A594}, 460 (1995).
\bibitem{e895}
P. Chung {\it et al.}, Phys. Rev. Lett. {\bf 85}, 940 (2000).
\bibitem{pal}S. Pal, C.M. Ko, Z.W. Lin, and B. Zhang, Phys. Rev. 
C {\bf 62}, 061903 (2000).
\bibitem{kaos}C. Sturm {\it et al.}, Phys. Rev. Lett. {\bf 88}, 39 (2001).
\bibitem{fuchs}C. Fuchs {\it et al.},
Phys. Rev. Lett. {\bf 86}, 1974 (2001).
\bibitem{rate}C.M. Ko, V. Koch, Z.W. Lin, K. Redlich, M. Stephanov, and 
X. N. Wang, Phys. Rev. Lett., in press; nucl-th/0010004.
\bibitem{kyield}R. Bath {\it et al.}, 
Phys. Rev. Lett. {\bf 78}, 4027 (1997).
\bibitem{transport}C.M. Ko and G.Q. Li, J. Phys. G {\bf 22}, 1673 (1996);
W. Cassing and E.L. Bratkovskaya, Phys. Rep. {\bf 308}, 65 (1999);
S. Bass {\it et al.}, Prog. Part. Nucl. Phys. {\bf 42}, 313 (1999).
\bibitem{kaon}J. Randrup and C.M. Ko, Nucl. Phys. {\bf A343}, 519 (1980);
J. Aichelin and C.M. Ko, Phys. Rev. Lett. {\bf 55}, 2661 (1985); 
X.S. Fang, C.M. Ko, and Y.M. Zheng, Nucl. Phys. {\bf A556},
499 (1993); X.S. Fang, C.M. Ko, G.Q. Li, and Y.M. Zheng, Phys. Rev.
C {\bf 49}, 1139 (1994); Nucl. Phys. {\bf A575}, 766 (1994); 
G.Q. Li and C.M. Ko, Phys. Lett. B {\bf 349}, 405 (1995).  
\bibitem{kaon1}S.W. Huang {\it et al.},
Phys. Lett. B {\bf 298}, 41 (1993); T. Maruyama {\it et al.},
Nucl. Phys. {\bf A573}, 653 (1994); C. Hartnack {\it et al.},
Nucl. Phys. {\bf A580}, 643 (1994).  
\bibitem{llb}G.Q. Li, C.H. Lee, and G.E. Brown, Nucl. Phys. {\bf A625},
372 (1997).
\bibitem{rvuu}C.M. Ko, Q. Li, and R. Wang, Phys. Rev. Lett. {\bf 59}, 1084
(1987); C.M. Ko, and Q. Li, Phys. Rev. C {\bf 37}, 2270 (1988); Q. Li, 
J.Q. Wu, and C.M. Ko, Phys. Rev. C {\bf 39}, 849 (1989); C.M. Ko, 
Nucl. Phys. {\bf A495}, 321c (1989).
\bibitem{kcross}G.Q. Li and C.M. Ko, Nucl. Phys. {\bf A594}, 439 (1995).
\bibitem{tsushima}K. Tsushima, S.W. Huang, and A. Faessler, Phys. 
Lett. B {\bf 337}, 245 (1994); J. Phys. G {\bf 21}, 33 (1995).
\bibitem{kaplan}D.B. Kaplan and A.E. Nelson, Phys. Lett. B 
{\bf 175}, 57 (1986).
\bibitem{sibirtsev}A. Sibirtsev, W. Cassing, and C.M. Ko, Z. Phys. A 
{\bf 358}, 101 (1997).
\bibitem{cugnon}J. Cugnon, P. Deneye, and J. Vandermeulen, Phys. Rev. 
C {\bf 41}, 1701 (1990).
\bibitem{ko1}C.M. Ko, Phys. Lett. B {\bf 120}, 294 (1983); {\it ibid.} 
{\bf 138}, 361 (1984).
\bibitem{akaon}G.Q. Li, C.M. Ko, and X.S. Fang, Phys. Lett. B {\bf 329},
149 (1994).
\bibitem{cassing}W. Cassing {\it et al.},
Nucl. Phys.  {\bf A614}, 415 (1997).
\end{thebibliography}
\end{document}